\documentclass{article}

\usepackage{PRIMEarxiv}

\usepackage[utf8]{inputenc} 
\usepackage[T1]{fontenc}    
\usepackage{hyperref}       
\usepackage{url}            
\usepackage{booktabs}       
\usepackage{amsfonts}       
\usepackage{nicefrac}       
\usepackage{microtype}      
\usepackage{lipsum}
\usepackage{fancyhdr}       
\usepackage{graphicx}       
\graphicspath{{media/}}     

\title{GraphLED: A graph-based approach to process and visualise linked engineering documents
}

\author{
  Vanessa T. da Silva, Lucas de A. M. Ribeiro, Willian B. de Lemos, \\ \textbf{Sílvia S. da C. Botelho}, \textbf{Nelson L. D. Filho}, \textbf{Marcelo R. Pias} \\
  Universidade Feral do Rio Grande (FURG) \\
  Rio Grande, RS, Brazil\\
  \texttt{\{vanessa.telles, lucasribeiro, willianlemos.b, silviacb, dmtnldf, mpias\}@furg.br} \\
}

\begin{document}
\maketitle

\begin{abstract}
The architecture, engineering and construction (AEC) sector extensively uses documents supporting product and process development. As part of this, organisations should handle big data of hundreds, or even thousands, of technical documents strongly linked together, including CAD design of industrial plants, equipment purchase orders, quality certificates, and part material analysis. However, analysing such records is daunting for users because it gets complicated to sift through hundreds of documents to establish valuable relationships. This paper addresses how knowledge extracted from linked engineering documents contributes to industrial digitalisation under IT/OT convergence. The proposed GraphLED is a system tasked with data processing, graph-based modelling, and colourful visualisation of related documents. The graph-based approach ensures an improved understanding of linked information because the graph structure offers a promising tool to model the underlying data properties of engineering documents. Preliminary system validation indicates quality improvements are possible in the OCR-based data  (85.9\% of ambiguous text data removed). This work has the potential to benefit the industry by improving the reliability and resilience of industrial production systems through automated summaries of large quantities of documents and their linkage.
\end{abstract}



\section{Introduction}
\label{sec:intro}

Ubiquitous data creation has grown at a rapid pace, with big data collection providing tangible digital transformation benefits in several industries \cite{wollschlaeger2017future}. However,  data remains document-centric in mixed analogue, and digital formats in more traditional sectors, including architecture, engineering and construction (AEC). Organisations often deal with hundreds, or even thousands, of technical documents such as CAD design of industrial factory plants, fleet design, equipment purchase orders, quality certificates, and material analysis  \cite{amor1997documents}. 

The management of technical documents with efficient tools leads to efficient coordination of critical day-to-day activities throughout AEC process lifetime \cite{zheng2019towards}.To this end, continuous data processing and visualisation bring considerable benefits to the AEC end-users: (i) support timely inspection of engineering projects, (ii)  develop automated document compliance models with existing standards and societies' guidelines and (iii) promote information exchange among key stakeholders (suppliers, clients, standards societies). However, AEC documents are inherently complex because diverse unstructured topics usually link many documents to convey the information. Conventional state-of-the-art approaches that handle structured data (e.g. search by keywords in relational databases) are not feasible in this case \cite{matsubayashi2018research}. 

The relationship between the tables is defined in advance, so  users can only obtain limited relevant knowledge \cite{guan2019application}. Paper-based documents further complicate the matter where scanned handwritten reports, technical specifications, engineering drawings, schematics, stamps, signatures and so on \cite{dahl2019document} make it challenging to retrieve information efficiently, even in its simplest form. For instance, a \textit{databook} --a collection of engineering documents that specify project events -- comprises many pieces of data (i.e. topics) that coherently link with data present in other databooks. This interconnection creates a complex yet powerful \textit{network} of related engineering documents that offer several features. For instance, the end-users can visually inspect the material traceability of equipment parts and their compliance with current technical specifications (e.g. ASTM - American Society for Testing and Materials). Today the visualisation and inspection of a large dataset of documents become challenging and error-prone if the user is not assisted by big data tools.  

A key high-level question in this context is: \textit{how could knowledge, namely engineering document summaries and their relationships, extracted from noisy OCR processed engineering documents contribute to industrial digitalisation and resilience of production systems?}  

The knowledge extraction is expected to be carried out efficiently and close to real-time. Equally important in this process is the exploitation of the cross-reference data between documents to obtain new information, a task that is inherently difficult for humans because of the document network depth and broadness  \cite{hovy2013collaboratively}.

This paper sheds some light to answer the question above. The proposed approach considers a simple yet promising  mechanism to extract and structure the data for processing, knowledge extraction, and visualisation from diverse engineering databooks. The GraphLED is a system tasked with data processing, graph-based modelling, and visualisation of related documents. The graph-based approach ensures an improved understanding of linked information because the graph structure offers a first-hand tool to represent and model the underlying data properties \cite{10.1007/978-3-030-14401-2_29}. 

The system assumes a moderate level of correctness in the close-to-real-time knowledge extraction from scanned documents, particularly cross-reference data. For instance, ambiguous data from OCR processing introduces errors that should be handled accordingly. 

This paper is structured as follows: Section 2 discusses related work. Section 3 explores user requirements and GraphLED design choices. The system architecture is discussed in Section 4. Finally, sections 5 and 6 present the validation results and draw the paper's conclusions.

\section{Related Work}
\label{sec:related}

Engineering databooks are complex documents. The user's acceptable level of data correctness associated with the knowledge extraction remains an open challenge, particularly considering the output quality obtained from the available OCR tools (e.g. Tesseract, Google OCR). The issue is beyond the OCR tools, where low-quality scanning of documents remains the norm in the industry. Conventional approaches that handle structured data are not very robust in this case, as noisy data in document table cells lead to limited relevant 
knowledge \cite{matsubayashi2018research} \cite{guan2019application}. 

Several applications use graph-based analysis to extract knowledge and visualise the results. For example, \cite{lee2015social} seeks to create public awareness of UN members voting patterns. The approach uses social knowledge graphs and similarity measures to analyse the publicly available information on the member votes. Because graphs lend themselves to intuitive visualisation, this work explored visual aids from the data. However, such an approach is less robust to data noise since stages of acquisition, treatment, and handling of the data require a prior pre-processed and ready-for-use dataset. Linked engineering AEC documents establish connections (e.g. purchase order number, quality certificate ID) that can be followed and analysed through automated software tools \cite{eto2019extended}. 

The graph-based analysis explores algorithms on graph data structures  \cite{strohman2007recommending}. For instance, the graph-based document representation system (CGDR) provides classification reports intended for document forensic analysis \cite{mujtaba2018classification}. A total of 1500 files were used in the study to classify 16 types of reports. The method structures the information as graph nodes and uses hierarchical text classification to set the reading order. However, the proposed model is not the ideal solution as it still depends on human experts to identify the useful features for each type of report. 

A more automated and adaptive approach could benefit the application in many ways. Also, the data should be inputted in an advanced processing stage (e.g. augmented, cleaned) for the system to work. The work in \cite{zhang2020every} proposes a graph-based method for inductive text classification in documents. 

The system builds individual graphs for each document fed into a Graph Neural Network (GNN) that learns fine-grained word representations based on their local structures. The proposed approach can effectively produce embeddings for unseen words in new documents. However, the noisy data limit the robustness of the proposed work. \cite{rasmussen2019managing} introduces knowledge graphs and ontologies to cope with the complex network of decisions in construction documents. Other related work explores methods and algorithms in automated analysis of engineering drawings, commonly used in the oil and gas engineering industry. 

Authors in \cite{8893107} acknowledge the document quality as being the backbone of the problem, given that a poor-quality drawing (e.g. a paper that has been scanned numerous times) is virtually impossible to digitise unless a robust set of pre-processing methods are used. 

The proposed work addressed the problem using a deep neural network with satisfactory improvements. However, the approach remains model-centric (i.e. model hyperparameter tuning) with little effort in improving the raw data quality through pre-processing, cleaning, augmentation or re-shaping. Existing work explores to a great extent digitalisation of documents through OCR (Optical character recognition) based techniques \cite{moreno2019new} and \cite{moreno2016graph}. The benefits of the links between documents to unveil very relevant patterns in knowledge extraction are not sufficiently explored in the previous work.

The prior art acknowledges the challenge of achieving a satisfactory level of data correctness regarding close-to-real-time knowledge extraction from scanned documents, particularly cross-reference data. Data noise propagates throughout the knowledge extraction processing pipeline to visualisation and user engagement. 

The previous work primarily addresses the correctness problem using human-intensive correction  \cite{matsubayashi2018research} \cite{guan2019application} or, more recently, using model-centric deep neural networks \cite{8893107}, where the job is to fine-tune model hyperparameters as opposed to improve input data quality. 
This paper advances the state-of-the-art as follows: 
\begin{itemize}
\item Proposes and validates data cleaning, and augmentation techniques (i.e. data-centric methodology) in noisy OCR scanned documents to overcome the so-called \textit{garbage in, garbage out} error propagation.
\item Proposes a system architecture to provide a large scale graph-based tool for close-to-real-time knowledge extraction and representation, assisting users in the difficult task of sifting through hundreds to thousand of documents.  The system validation focuses on the performance and scalability issues.
\end{itemize}

\section{System Design Choices}

The proposed work is part of a larger technology-driven project that aims at digitalising industrial documents (OCR-based) grouped in databooks units. The work realises the vision of building digital twins tailored for the Oil\&Gas industry to benefit users whose roles include inspection of documentation of materials, equipment, construction processes and quality control as part of the IT/OT convergence trend. A series of meetings and workshops were held with AEC documentation inspectors in 2021 that elicited the following list of user requirements:

\begin{itemize}
    \item Ability to upload and extract key information from a large quantity of scanned databooks through OCR-based systems. The expected correctness level is minimal ambiguity in the knowledge-extracted data.  
    \item Capability to visualise, manipulate and analyse databooks through intuitive and easy-to-use interfaces.
    \item Carry out usual inspection tasks (Completeness and Conformance) in an automated fashion. 
    \item Assert that specific information extracted from a databook complies with existing technical specifications and standards (e.g. ASTM B16). 
    \item Discover where and how the product and process were made (traceability, anomaly detection). 
\end{itemize}

\subsection{Databooks}
The data space of interest comprises technical documents that make references to other documents. For instance, an equipment purchase order document can have relationships with complementary documents describing the chemical and physical properties of the materials used. 

Another example is the collection of information that documents the construction process of offshore oil rigs and platforms. In this context, a databook is a group of related documents covering relevant aspects of a specific product or process cycle, from material production, batch production, product conformance validation and purchase order description. 

For instance, the purchase of a bolt for a high-pressure gas pipe is described in a databook. A series of documents specify the bolt part's key aspects, including the manufacturing process, material properties, ultra-sound testing, and visual, and dimensional characteristics. 

\begin{figure}[ht]
\centering
\includegraphics[width=.4\textwidth]{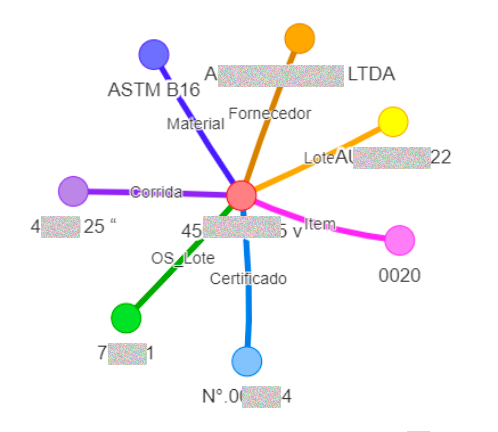}
\caption{Complete Graph: documents linked through topics (e.g. batch number OS\_LOTE). }
\label{fig:graforcCompleto}
\end{figure}

\begin{figure}[ht]
\centering
\includegraphics[width=.4\textwidth]{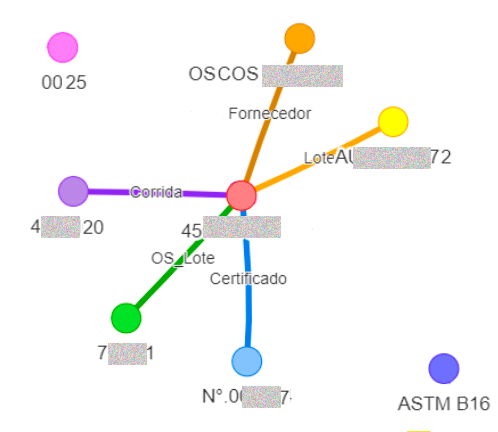}
\caption{Incomplete Graph: databook with missing links. }
\label{fig:graforcincompleto}
\end{figure}

\subsection{Design Choices}
The system should check the completeness and compliance of databooks with relevant standards, databook traceability information and anomaly detection in case of data inconsistency and potential fraud. Non-functional requirements include low-latency operation, real-time performance, data coherence and integrity.

\subsubsection{Completeness and conformance}
Inspecting completeness comprises checking whether all required documents are readily available in the databook. In addition, conformance to standards and guidelines is carried out to check several compliance rules such as material properties thresholds, dimensional specifications and visual inspection.

\subsubsection{Traceability}
This system functionality ensures that a product or process has records that prove the manufacturing path to its inception. The traceability inspection searches all databook documents referenced in the dataset to analyse the linked and source data compatibility. 
In addition, incomplete databooks or non-conformant documents should also be flagged as faults in the traceability verification. 

\subsubsection{Data Coherence}
Data inconsistency is inevitable as the documents are scanned, OCR (Optical Character Recognition)  processed to extract relevant text, and structured in an automated way. For instance, word ambiguities, either because of language variability or OCR quality issues, challenge the proposed system where the same data element has different meanings. In this case, the system should strive to maintain consistency and coherence at all costs. 

\subsubsection{Linked Documents}
Graph-based structures and models provide a design choice to specify data elements and relationships of technical documents. Also, several graph algorithms can be explored to fulfil the document verification needs, including completeness, conformance and traceability. For instance, centrality techniques traverse a graph to compute such properties as popular elements and bottlenecks. 

The visual appeal of graph-based data representation offers a more intuitive data exploration to document inspectors. Figs. \ref{fig:graforcCompleto} and \ref{fig:graforcincompleto} shows an example of a databook sample, represented as a graph, extracted from a real-world dataset (Oil\&Gas industry, identifiers have been crossed-out). In this case, a graph $G(V, E)$ comprises a set of vertices $V$ (databook individual documents) and a set of edges $E$ that interconnects the vertices. Fig. \ref{fig:graforcCompleto} shows an example of a graph with a complete databook, given that there is a path connecting all documents through the central node. In contrast, Fig. \ref{fig:graforcincompleto} presents an incomplete graph where documents are missing, and two documents are isolated, thus compromising the intra-databook traceability. It is important to note that these documents relate to other documents through specific relationships (e.g. same quality assurance laboratory).

\section{The GraphLED System}
\begin{figure}[ht]
\centering
\includegraphics[width=.7\textwidth]{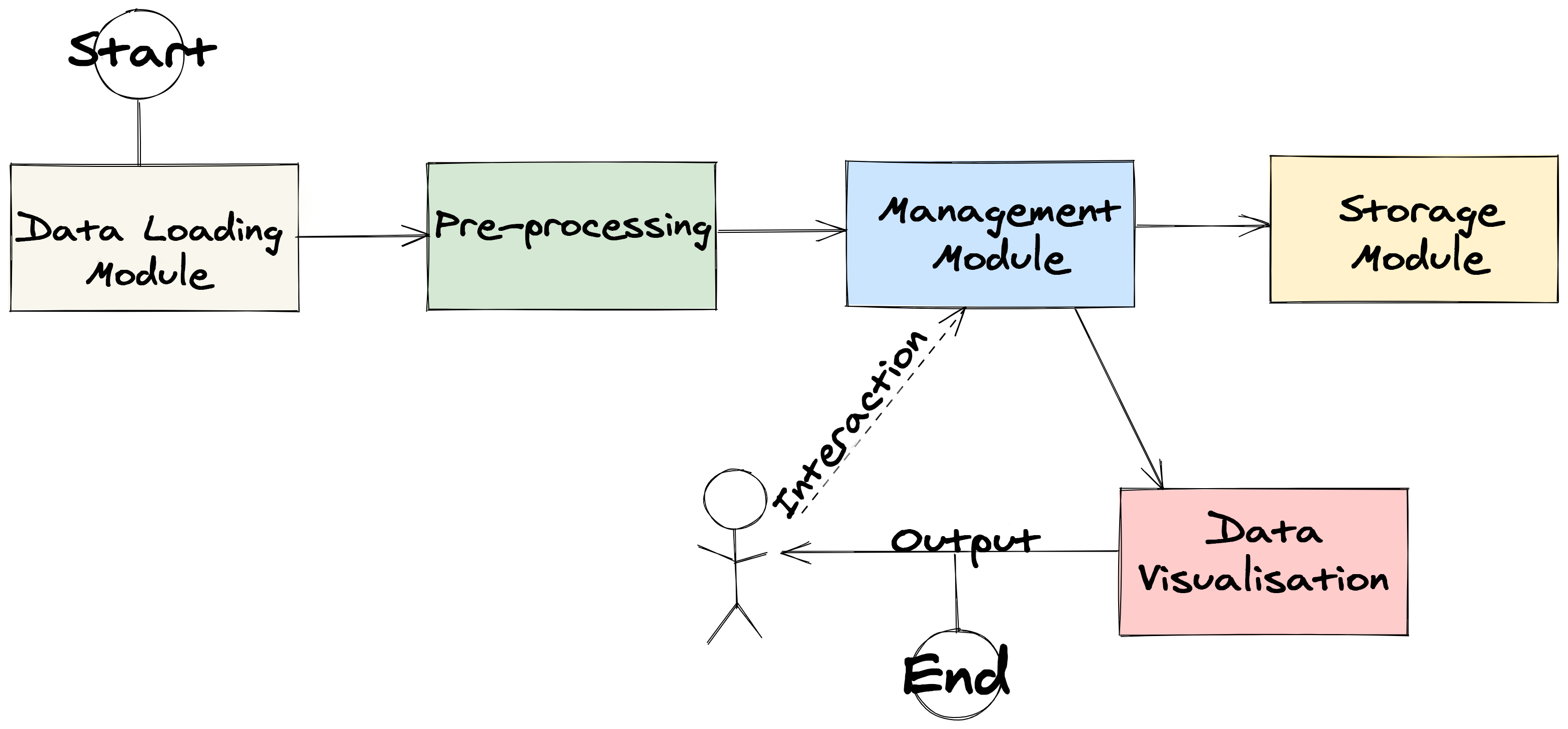}
\caption{System architecture modules. The GraphLED explores a colourful visual representation using bright colours (i.e. analogy to LEDs)}
\label{fig:modulos}
\end{figure}

The GraphLED system architecture (as shown in Fig.~\ref{fig:modulos}) is modularised to provide interoperability in the implementation of the pipeline functions from data insertion and knowledge extraction to graph visualisation and processing.

\subsection{Data Loading Module}
OCR tools (e.g. Tesseract, Google image reader) convert scanned documents into raw text under a certain quality expectation. To minimise the OCR-related errors, GraphLED adopts the steps proposed in \cite{santos2021form}, where the generated raw text data is pre-processed using noise reduction, binarisation and skew reduction.

The outputted document is fed into a \textit{form understanding} process that classifies documents using a form-based layout similarity scheme of document models previously defined. This module output is a JSON file containing each document represented as a set of blocks with the extracted words, their bounding boxes, and whether there is a link to a piece of data in another document.


\subsection{Pre-Processing}
Data is inherently noisy in this application scenario. This module handles the data cleaning and wrangling to improve data quality by transforming the input \textit{dataset} into an intermediate representation suitable for the graph generation. Pre-processing tasks are performed at coarse-grained (documents) and fine-grained (document words) levels. A summary of the challenges is as follows: (1) Data ambiguity: Data is represented differently, but it semantically represents the same information; (2) Irrelevant data: data with small intrinsic value for a graph-based analysis and (3) Over-relevant data: identifying the more likely data to become nodes or edges in the graph generation phase. 

A sequence of operations is applied to the data to decrease noise levels. For instance, it is crucial to filter out \textit{Stopwords} prior to using disambiguation functions. Disambiguation works through similarity comparisons between data instances, using the following filters: 
\begin{itemize}
    \item Levenshtein filter \cite{haldar2011levenshtein} relies on the edit distance between two strings of comparable size (same proportion) to minimise false positives. A low acceptance threshold is also used to detect very similar instances.
    \item The Longest Common Subsequence (LCS) filter \cite{tiedemann1999automatic} detects similarity through the longest common subsequence (in the exact relative order) between any two strings.
    \item Sequence Matcher filter (LCS variation) uses junk-free contiguous subsequences as a similarity metric. 
\end{itemize}

\begin{figure}[ht]
\centering
\includegraphics[width=.7\textwidth]{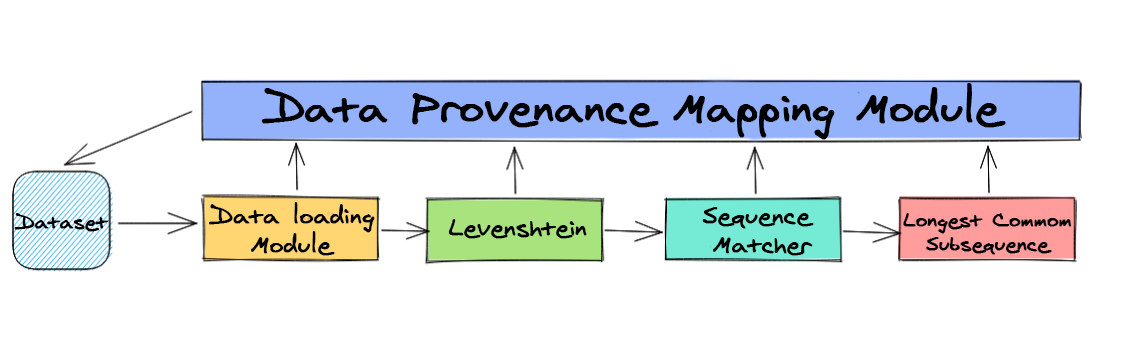}
\caption{Proposed ambiguity module architecture. }
\label{fig:FiltroDiagrama}
\end{figure}

Fig.\ref{fig:FiltroDiagrama} presents the pipeline for the ambiguity correction sub-module, where filters are applied one after another, covering the primary forms of ambiguity found in the dataset. Another module maps the modifications performed in the filtering steps to instances of the original document for data provenance purposes. It is essential to keep track of input data, transformed data and processes, thus providing a historical record of the data and its origins.

With a large amount of data, exploiting helpful information can be daunting. Nevertheless, once we decrease the noise with filtered data, graph-based algorithms can attempt to spot patterns to detect relevant information based on past data behaviours. The detection of relevant data nodes exploits graph centrality algorithms for trend identification. Such algorithms enable the monitoring of a set of indicators: (a) the transitive influence of nodes (Eigenvector Centrality), (b) node influence on the flow of the graph (Betweenness Centrality) and (c) the number of incoming and outgoing relationships from a node (Degree Centrality).

The criterion for defining the relevant features is the aggregate sum of the scores produced by the metrics above.

\subsection{Management Module}
\begin{figure}[ht]
\centering
\includegraphics[width=.6\textwidth]{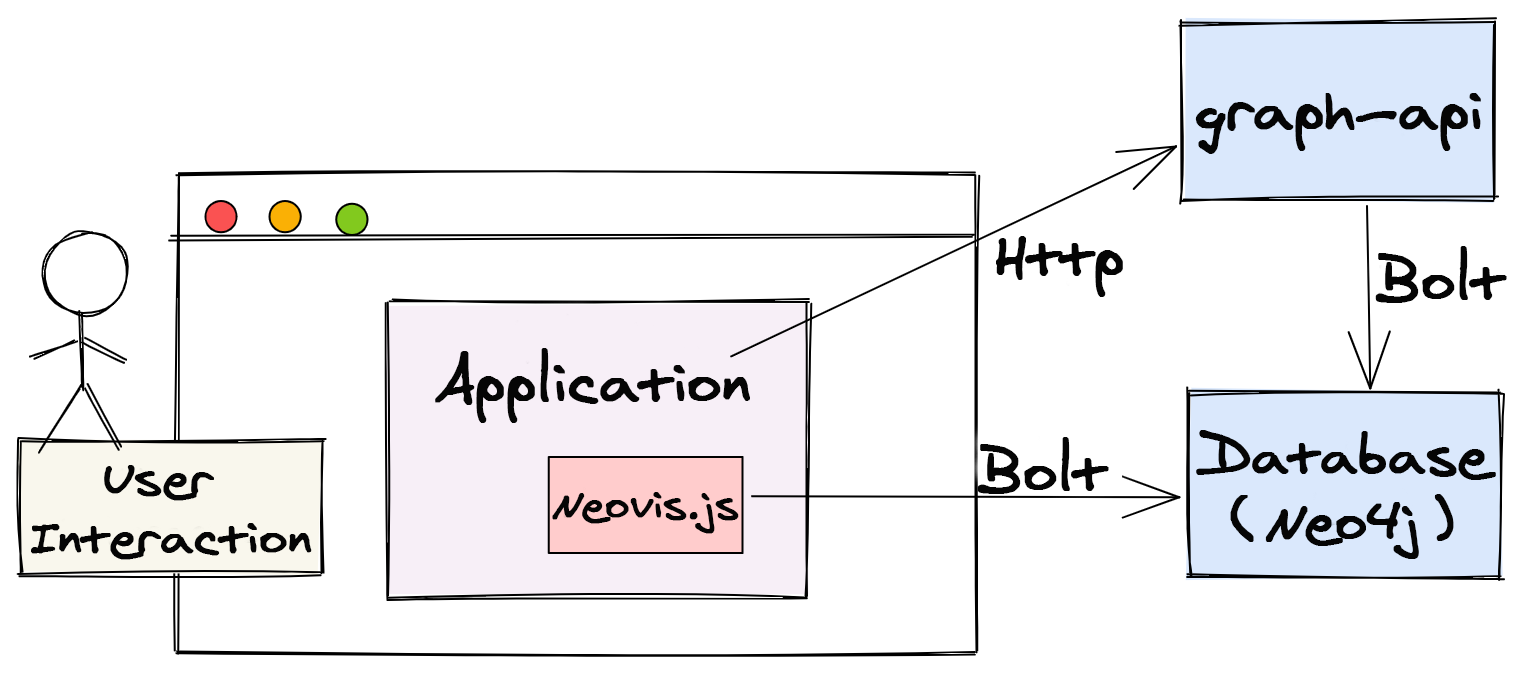}
\caption{Management Module. Data is exchanged with the Neo4j graph database through the Bolt protocol 
}
\label{fig:gerenciamento}
\end{figure}

This module deals with system integration and its proper operation. It exchanges information through RestFUL APIs (Fig.~\ref{fig:gerenciamento}), namely  \textit{graph-api}, that allows \textit{query} creation considering the user parameters informed via HTTP requests. 

\subsection{Storage Module}
Graph-based storage and query engine are essential functions in the GraphLED system. The following criteria were used for the selection of a suitable graph database system:

\begin{itemize}
    \item Maturity and availability of documentation and support from the developer team and community. 
    \item Easy integration with other existing tools and provision of programming language integration (e.g. Python, Java and JavaScript). 
    \item Satisfactory performance in moderate data volume.  
    \item Easy-to-use, understandable and maintainable graph query language. 
    \item Schema-free: flexibility and agility in structuring graphs. 
\end{itemize}

Unlike traditional databases (SQL-like), graph-oriented databases offer a few advantages, but the primary one is the optimisation of breadth-first searches, which is a highly recommended procedure to find non-obvious relationships between entities. A few choices were considered, including cloud-based Amazon Neptune, Azure CosmoDB and community-based Neo4j. The latter fulfilled the project requirements at the time. Also, the upcoming query language standards (Apache Gremlin and SparkQL) can soon make the GraphLED system graph database-agnostic.

\subsection{Data Visualisation}
Graphs lend themselves to intuitive data visualisation. This module uses open-source toolkits, particularly the Neovis library, to render graph nodes and edges that model databooks, associated documents, and their relationships. In addition, the GraphLED visual tool interacts directly with the Bolt protocol's graph database for improved performance. However, it has been noted that graph visualisation can be a bottleneck as the number of graph nodes increases. 

The designed user interfaces mitigate potential bottlenecks through \textit{(i)} simple searches aimed at node-edge-node traversals, in which the user can specify the nodes' classes. 

The system offers \textit{(ii)} advanced \textit{queries} in  \textit{Cypher} language, returning to the user complex and elaborated results. Properties calculated \textit{(iii)}, such as degree centrality, are also presented in a table format at the bottom of the GUI.

\subsection{User Interface}
\begin{figure}[ht]
\centering
\includegraphics[width=.8\textwidth]{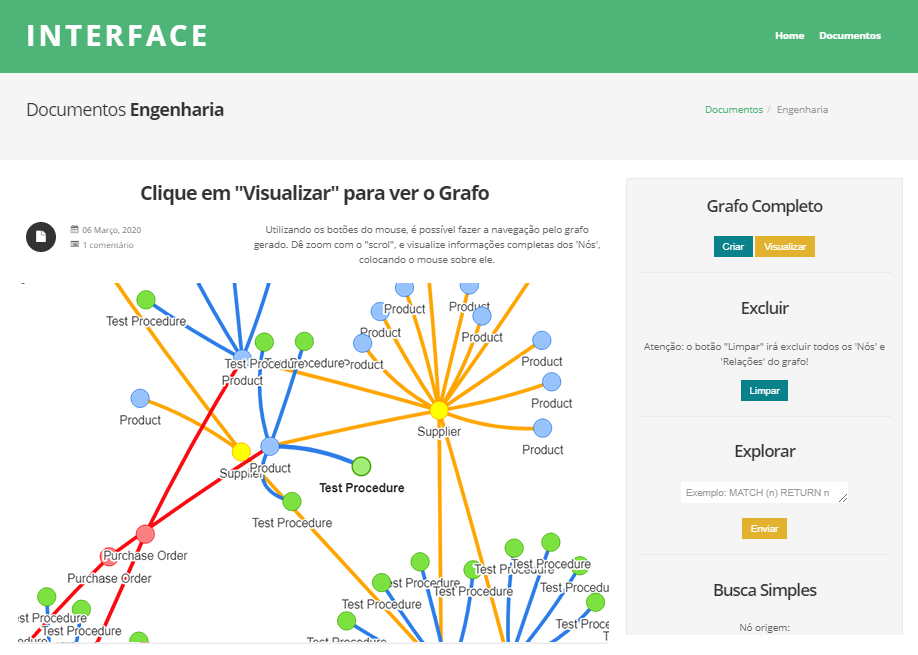}
\caption{Beta version. }
\label{fig:interface}
\end{figure}

The system \textit{beta} version (Fig. \ref{fig:interface}) offers aesthetic and improved user experience. Customised GUI elements were added, allowing users to choose between the various types of engineering documents, and create, visualise and delete databook/document graphs. 

An exploration form field has been placed in the side menu so users can write complex graph traversal searches, choosing between nodes and connections types.  In addition, a simple search box was added to complement node centrality information, where users can choose the following properties: \textit{Betweenness, Node Degree, Closeness and Eigenvector centrality.} The centrality provides hints to the user on structural anomalies, such as incomplete graphs, missing relationships, and others. Given that databook's documents are modelled as graphs, the centrality properties answer the questions of completeness, conformance, and traceability.

\section{System Validation Results}
\subsection{Data Quality}

The validation is aimed at understanding and further improving the input data quality. 

The data pipeline implemented takes in a scanned document and passes it through an OCR stage. Then the OCR outputted file is classified and stored. The validation dataset used 81 PDF scanned databook files, totalling \textit{1.43 GB} of individual documents. It is important to note that a databook is a set of engineering-driven documents with various purposes ranging from purchase order specifications to material testing certificates. 

The benchmarking employed two document categories, labelled as \textit{easy document} and \textit{difficult document}. The labels indicate how difficult it is for the Tesseract OCR system to process scanned documents considering visual and document structural aspects such as page inclination, scrawls and complex internal elements (e.g. tables with few lines, few pages).

Further validation experiments were undertaken on a small set of labelled data to analyze the performance of data noise filtering and disambiguation techniques (Pre-processing module). In addition, many recognized data entities in the graph have been used as a comparison metric because it conveys the amount of duplicate/ambiguous nodes in the dataset. The data comprises parts suppliers to oil platforms (the dataset has 226 rows). Specialists analyzed the dataset and found 17 distinct suppliers (out of 226). However, the graph generated where each supplier is a node contained 128 different suppliers.

\begin{figure}[ht]
\centering
\includegraphics[width=.45\textwidth]{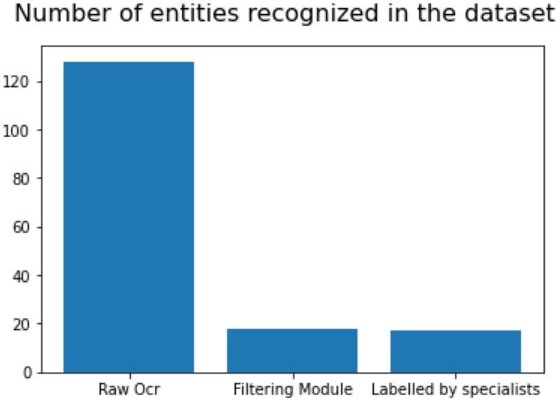}
\caption{Comparison between recognized entities by the OCR system, after the filter, and by specialists.}
\label{fig:entities-recognized}
\end{figure}

Further analysis found that such a discrepancy was due to ambiguous nodes in the graph, caused by 128 grammatical variations of the 17 suppliers. Some examples of variations include the terms: "supplier-A": "A-supplier", "supplier-A.inc", "group-A-supplier". The key idea for dealing with this issue is to identify variations of the same entity (a supplier) in the dataset and standardize them in a single form of writing. The ambiguity removal algorithm was developed to accomplish this task. The algorithm produced a list of 18 distinct suppliers, just one supplier off the specialist's list,  as is summarised in Figure \ref{fig:entities-recognized}. As a result, the algorithm removed 110 ambiguous nodes in the graph - total removal of 99.09\% of ambiguous nodes and an 85.93\% reduction in the number of nodes from 128 to 18 suppliers.

\begin{figure}
\centering
\includegraphics[width=.47\textwidth]{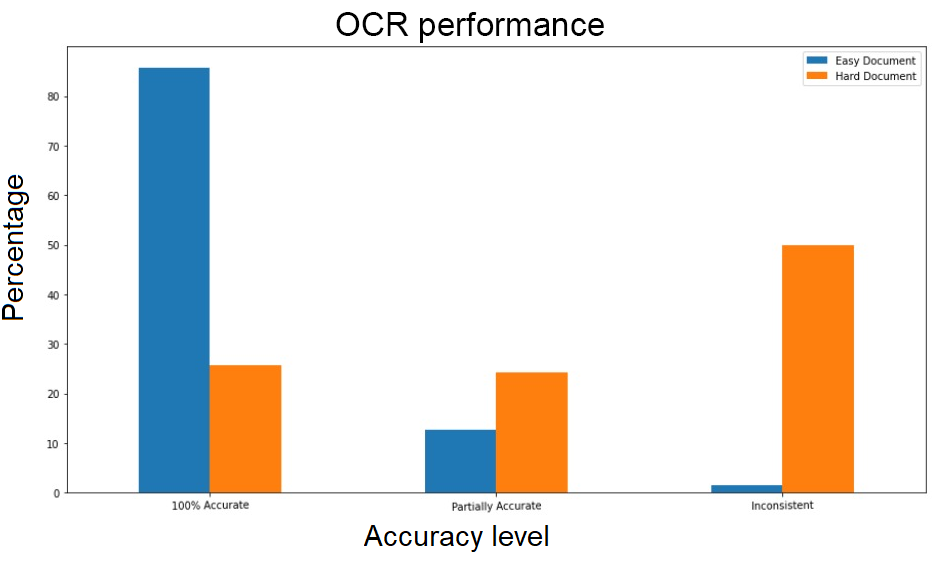}
\caption{OCR performance: easy and difficult documents. }
\label{fig:OCR-scenarios}
\end{figure}

The results were classified considering the similarity (OCR text versus ground-truth original text) into: \textit{100\% Accuracy} (total hit), \textit{Partially accurate} (partial hit) and \textit{Inconsistency} for lack of any relation between the OCR and ground-truth text. In the \textit{easy document} context, as shown in Fig. \ref{fig:OCR-scenarios}, the system managed to identify 85.9\% of the data elements inside the document with total hits, 12.72\% as partial hits and 1.58\% as inconsistent (total error). However, the results for the \textit{difficult document} were not satisfactory: 25.67\% of total hits, 24.32\% of partial hits and 50.01\% of inconsistencies. 

It is clear from these results that noise and ambiguity in the data propagate through the OCR-processed documents. Such errors are not helpful for graph construction from the OCR data. 

More complex forms of ambiguity, such as entity standardization and acronym identification, were handled manually and will be addressed in an automated way in the next release of the GraphLED system. 

\subsection{Scaling the GraphLED}

As a quantitative reference, an analysis of the Read/Write rate of graph nodes, edges and properties was carried out in the Neo4j database. We chose the \textit{graph-workload} \cite{davidallen19}, an open-source tool for running loading tests, which is capable of generating random loads of data and Cypher queries to create graph nodes, edges and properties in a configurable and scalable way.

Table~\ref{table:tabela1} lists the graph database setup parameters, including a Heap Initial Size value equal to 4GB of DDR4 RAM, limited by a Heap Max Size of 8GB.

The experiment parameters used are: \textit{n = 1000}, where ``n'' is the target number of queries in the database, and \textit{concurrency = 10} is the number of queries executed in parallel. The data used are generated randomly (at runtime) or deterministically (hard coded), having no relation with an input dataset. The selected subgraph patterns, e.g. the Hub-Spoke (Figure~\ref{fig:graforcCompleto}) and the N-Ary Tree pattern, are also observed in structures of operational industry documents.

\begin{table}[ht]
  \caption{Graph Database Setup Configuration.}
  \label{table:tabela1}
  \centering
  \begin{tabular}{ccl}
    \toprule
    Setting &\\
    \midrule
    Heap Initial Size & 4 GB\\
    Heap Max Size     & 8 GB\\
    Page Cache Size     & 4 GB\\
    Neo4j Version   & 3.5.14 \\
  \bottomrule
\end{tabular}
\end{table}

\begin{figure}[ht]
\centering
\includegraphics[width=.45\textwidth]{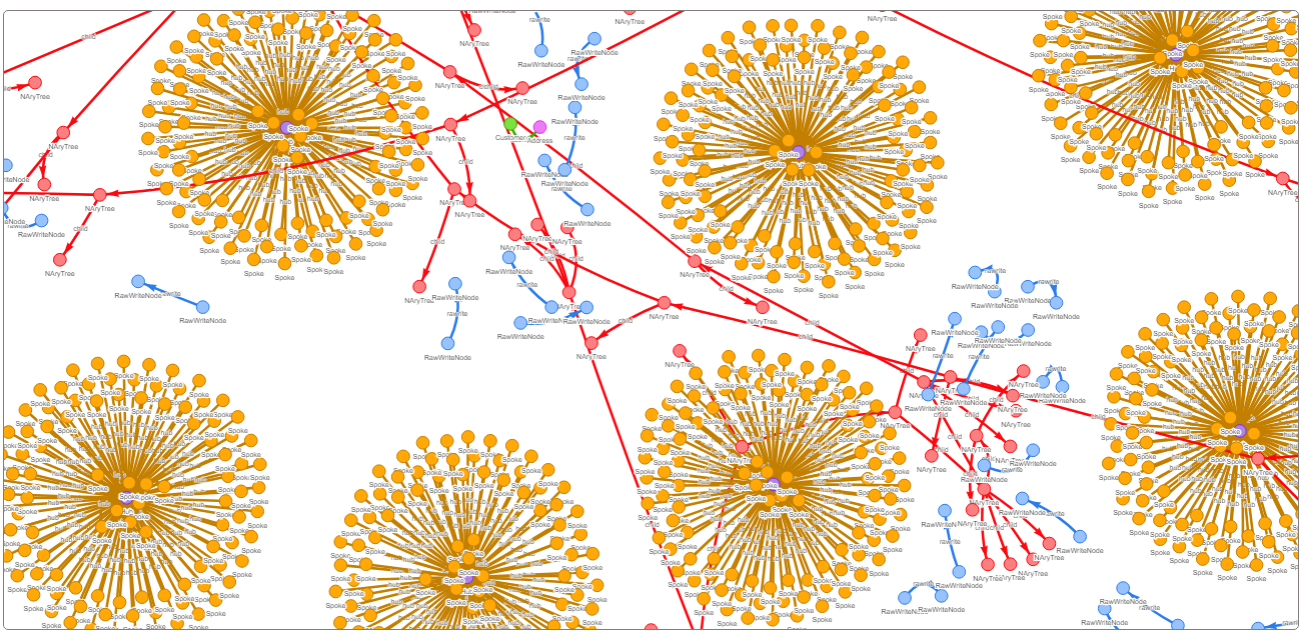}
\caption{Hub-Spoke Subgraph Pattern. }
\label{fig:benchmark2}
\end{figure}

\begin{figure}[ht]
\centering
\includegraphics[width=.45\textwidth]{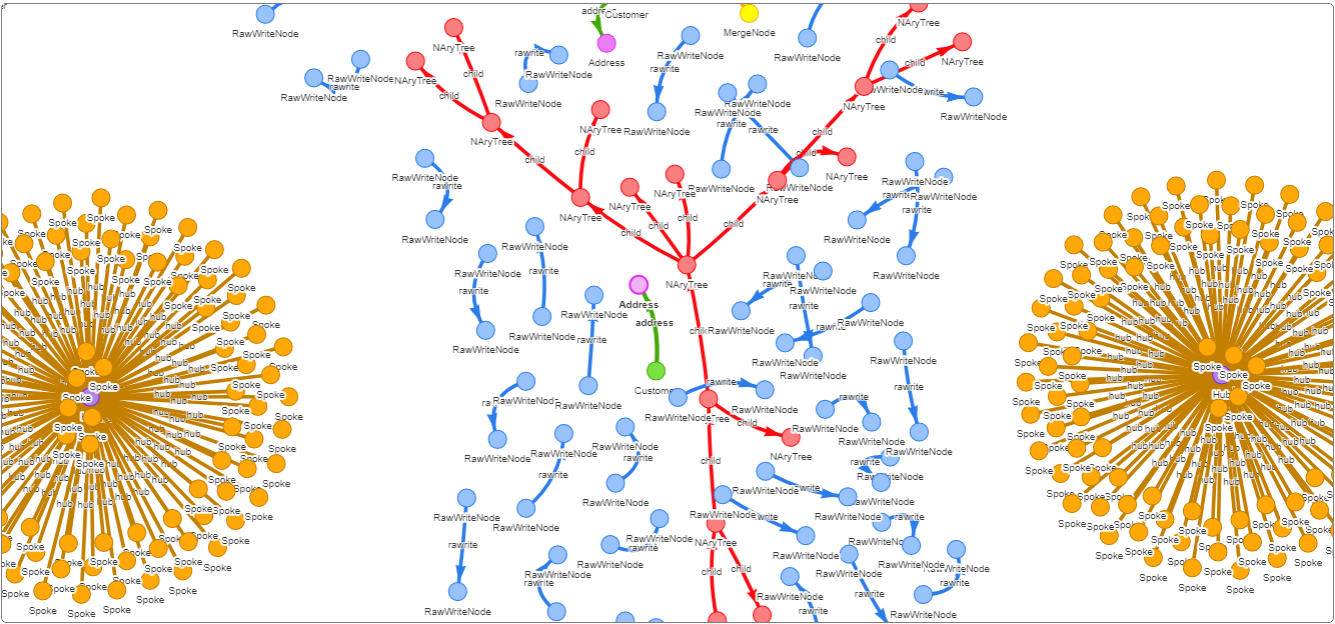}
\caption{N-Ary Tree Subgraph Pattern. }
\label{fig:benchmark3}
\end{figure}

\begin{figure}[ht]
\centering
\includegraphics[width=.45\textwidth]{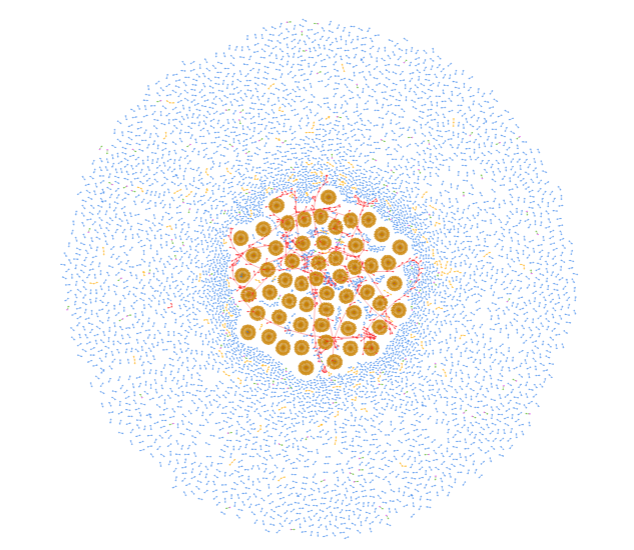}
\caption{Subgraph Sample with Ten Thousand Nodes. }
\label{fig:benchmark4}
\end{figure}

The complete graph could be generated with 143,602 nodes and 87,427 edges, with an approximate execution time of 42286 ms (Table~\ref{table:tabela2}). Figure~\ref{fig:benchmark4} shows the rendering of a graph sample limited to 10,000 nodes. We observe the following graph patterns and operations:

\begin{itemize}
    \item Hub-Spoke Structural Patterns can be spotted where a central node connects a "leaf" in a shape that resembles a dandelion. This type of node is represented in yellow (Figure~\ref{fig:benchmark2}).
    \item N-Ary Tree Pattern, with nodes in red representing an unbalanced N-Ary tree structure, where the large level of the sub-graph gives the main characteristic, are shown in red (Figure~\ref{fig:benchmark3}).
    \item Large set of 1-on-1 single linked structures to populate the graph where nodes are depicted in blue, pink and green in Figures~\ref{fig:benchmark3} and~\ref{fig:benchmark4}). In addition, some of such sub-graphs could exhibit other properties: a large number of indexable attributes (Index Heavy), sub-graphs of nodes with a large number of characters (Fat Node Append), sub-graphs with nodes connected through random links (Random Linkage), sub-graphs with both random nodes and links (Raw Write) and sub-graphs created from a MERGE operation.
    \item  Other types of operations (Aggregate Read, Long Path Read and Random Access Read) are allowed as Read queries directly performed to the database. Such operations were included only to simulate a production environment with real-world users when both Write and Read operations can coincide.
\end{itemize}

\begin{table}[ht]
  \centering
  \caption{Graph Database Running Results.}
  \label{table:tabela2}
  \begin{tabular}{ccccl}
    \toprule
    Pattern & RUNS & AVG & MIN & MAX\\
    \midrule
    Fat Node Append & 1050 & 73.03 & 18 & 1990\\
    NAry Tree & 1003 & 54.72 & 4 & 1993\\
    Merge Write & 985 & 38.24 & 4 & 1991\\
    Random Linkage & 1010 & 31.90 & 5 & 249\\
    Hub-Spoke & 481 & 60.74 & 20 & 719\\
    RawWrite & 2933 & 41.51 & 3 & 1754\\
    Index Heavy & 1002 & 41.10 & 4 & 1960 \\
    Aggregate Read & 552 & 18.25 & 2 & 126 \\
    Random Access Read & 952 & 37.44 & 7 & 114 \\
    Long Path Read & 62 & 24.66 & 2 & 127 \\
    \textbf{Total (units)} & \textbf{10030} & - & - & - \\
   
    \textbf{Average (ms)} & - & \textbf{42.16} & \textbf{6.9} & \textbf{1102.3} \\
  \bottomrule
\end{tabular}
\end{table}

\section{Conclusions}

The GraphLED system addresses the needs of Oil$\&$Gas inspection in \textit{big data} engineering contexts. 

This work contributes to a data-centric mechanism to address the correctness level in knowledge extracted from noisy OCR-based documents. The proposed data cleaning and augmentation achieved satisfactory results in removing ambiguity in the data. 

The GraphLED system offers a graph-based tool for close-to-real-time knowledge extraction and representation. 

\bibliographystyle{unsrt}  
\bibliography{references}

\end{document}